\documentclass[letterpaper, 10 pt, conference]{ieeeconf}  

\IEEEoverridecommandlockouts                              
\usepackage{graphics} 
\usepackage{epsfig} 
\usepackage{amsmath} 
\usepackage{amssymb}  

\usepackage{url}
\usepackage[ruled, vlined, linesnumbered]{algorithm2e}
\usepackage{verbatim} 
\usepackage{soul, color}
\usepackage{lmodern}
\usepackage{fancyhdr}
\usepackage[utf8]{inputenc}
\usepackage{fourier} 
\usepackage{array}
\usepackage{makecell}
\usepackage{tikz}
\SetNlSty{large}{}{:}
\usepackage{xcolor}

\usepackage{booktabs}

\makeatletter

\newcommand{\Rom}[1]{\expandafter\@slowromancap\romannumeral #1@}
\makeatother

\title{\LARGE \bf
Unmasking Dark Patterns: A Machine Learning Approach to Detecting Deceptive Design in E-commerce Websites
}

\author{Arya Ramteke, Sankalp Tembhurne, Gunesh Sonawane \& Prof. Ratnmala N. Bhimanpallewar  
\\ Department of Information Technology \\
 BRACT’s Vishwakarma Institute of Information Technology Pune, India. \\
{\tt\small arya.22110058@viit.ac.in, sankalp.22111127@viit.ac.in, gunesh.22111338@viit.ac.in \& ratnmalab@gmail.com}
}

\begin{document}

\maketitle
\thispagestyle{plain}
\pagestyle{plain}

\begin{abstract}

Dark patterns are deceptive user interfaces employed by e-commerce websites to manipulate users' behavior in a way that benefits the website, often unethically. This study investigates the detection of such dark patterns. These are some existing solutions to find dark patterns: UIGuard, a knowledge-driven system using computer vision techniques and natural language pattern matching to analyze UI elements and identify dark patterns by integrating taxonomies and distilling knowledge from real-world examples. Other approaches categorize dark patterns based on detectability into automated, manual, or undetectable, utilize explainable machine learning models trained on e-commerce dark pattern datasets, and explore generative AI technologies driven by regulatory interest. We propose a solution that combines web scraping techniques (BeautifulSoup4 and Selenium WebDriver) with contextual understanding through fine-tuned BERT language models and generative capabilities for identifying outlier dark patterns. Our approach involves scraping textual content from e-commerce websites, feeding it line-by-line into the fine-tuned BERT model for dark pattern detection, and leveraging BERT's bidirectional sentence analysis and generative abilities. The study builds upon such research on automatically detecting and explaining dark patterns in user interfaces, ultimately aiming to raise awareness and protect consumers from these deceptive practices.\\

\end{abstract}


\section{INTRODUCTION}

Dark patterns are manipulative design strategies often employed by e-commerce platforms to influence users' behavior and decision-making processes, often at the expense of ethical standards. These tactics exploit cognitive biases and utilize various methods to guide users towards actions that may not be in their best interests, such as making unintended purchases or sharing personal information. As the online shopping landscape continues to expand rapidly and regulatory scrutiny intensifies on consumer protection, detecting and addressing dark patterns has become increasingly critical.

This study delves into innovative approaches for automatically identifying dark patterns on e-commerce websites. Several existing solutions have been proposed, including UIGuard, a sophisticated system that employs computer vision techniques to analyze user interface elements and natural language pattern matching to detect dark patterns. By integrating established taxonomies and drawing insights from real-world examples, UIGuard offers a comprehensive framework for identifying these deceptive practices. Other methodologies categorize dark patterns based on their detectability, ranging from fully or partially automatable to those requiring manual inspection, with some remaining undetectable.

Moreover, the research explores the application of explainable machine learning models trained on datasets specifically curated to identify dark patterns in e-commerce settings. Generative AI technologies have also drawn attention due to regulatory efforts aimed at addressing deceptive online environments. In our proposed solution, we employ web scraping techniques using BeautifulSoup4 and Selenium WebDriver to extract textual content from e-commerce websites. This content undergoes line-by-line processing using fine-tuned BERT language models, facilitating contextual understanding and identification of dark patterns. Leveraging BERT's bidirectional sentence analysis and generative capabilities, our approach aims to effectively identify and explain potential dark patterns. By building upon existing research and methodologies, our solution aims to raise awareness and protect consumers from deceptive practices by providing a robust and scalable method for detecting and mitigating dark patterns on e-commerce platforms.
 
\section{OBJECTIVE}

Types of websites: The proposed solution aims to detect dark patterns across a wide range of e-commerce websites, encompassing various industries, product categories, and business models. This includes but is not limited to online retailers, marketplaces, subscription-based services, and any website that facilitates the sale of products or services to consumers.

Types of dark patterns to be detected: The solution will focus on identifying the following types of dark patterns commonly employed by e-commerce websites:
\begin{itemize}
\item Forced Action: Patterns that coerce users into taking specific actions, often by restricting their ability to proceed or complete a task unless they comply.
\item Misdirection: Patterns that intentionally misdirect or mislead users, obscuring important information or creating false assumptions.

\item Obstruction: Patterns that intentionally make certain processes or actions unnecessarily complicated for users.

\item Scarcity: Patterns that create a false sense of scarcity or urgency, pressuring users to make decisions quickly.

\item Sneaking: Patterns that attempt to sneak additional costs or unwanted subscriptions past users through deceptive means.

\item Social Proof: Patterns that leverage social proof or peer pressure to influence user behavior.

\item Urgency: Patterns that create a false sense of urgency or time pressure, prompting users to act quickly without careful consideration.

\item Not Dark Pattern: The solution should also be capable of identifying user interfaces or practices that do not constitute dark patterns, ensuring accurate and reliable detection.
\end{itemize}
The aim is to develop an all-encompassing solution capable of accurately detecting and categorizing various forms of dark patterns present on e-commerce platforms. This advanced tool will empower businesses, regulatory bodies, and consumers to effectively recognize and counteract these misleading tactics.

\section{METHODOLOGY}
\begin{figure}
    \centering
    \includegraphics[width=1\linewidth]{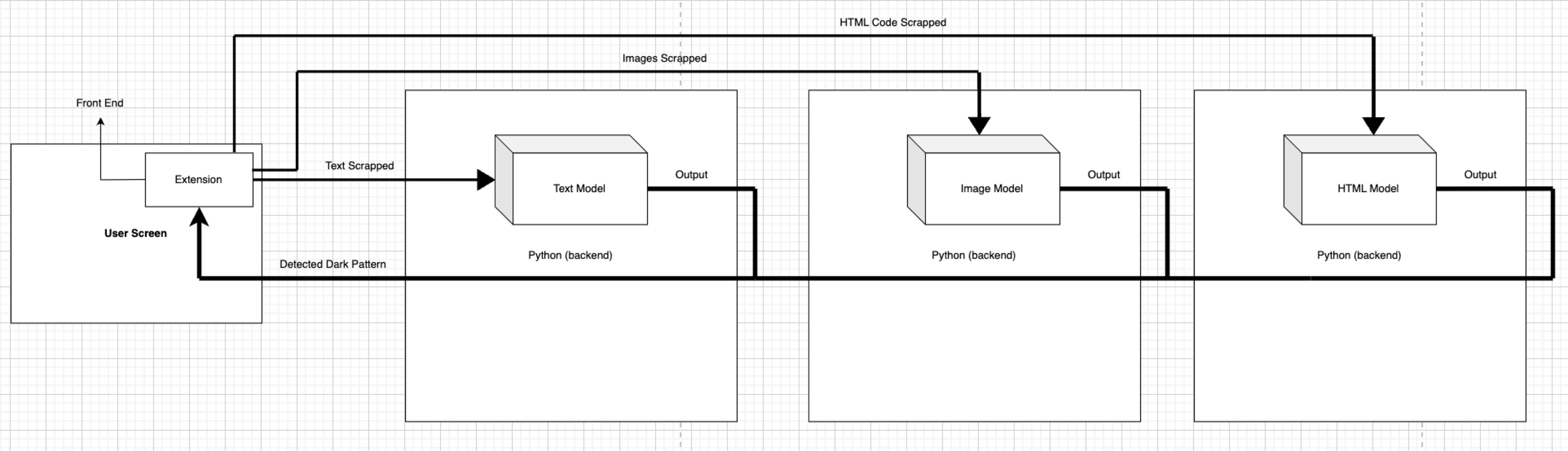}
    \caption{Pending Model Architecture}
\end{figure}
\subsection{Data Collection and Preprocessing}

researched various e-commerce sites and refined the data by human evaluators. Additionally, publicly available datasets from platforms like Kaggle and GitHub were utilized.

\subsection{Dark Pattern Annotation}

Human evaluators manually annotated the extracted text into eight categories: Forced Action, Misdirection, Obstruction, Scarcity, Sneaking, Social Proof, Urgency, and Not Dark Pattern, creating a refined dataset.

\subsection{Model Development and Training}

\textbf{BERT Fine-tuning:} The pre-trained BERT model served as the basis for our dark pattern detection system. We fine-tuned it on the annotated dataset through supervised learning.

\textbf{Hyperparameter Tuning:} Various hyperparameters were experimented with, including learning rate, batch size, and training epochs, optimizing the model's performance metrics like accuracy, precision, recall, and F1-score.

\textbf{Model Evaluation:} The fine-tuned BERT model achieved 96\% accuracy on the test set, demonstrating its efficacy in classifying text into different dark pattern categories.

\subsection{Dark Pattern Detection Implementation}

\textbf{Line-by-line Evaluation:} used Selenium WebDriver and BeautifulSoup (bs4) to scrape data from e-commerce websites and send it to the model for line-by-line evaluation.

\textbf{Probability Thresholds:} Thresholds for each dark pattern category were defined based on model performance and desired precision-recall balance.

\textbf{Dark Pattern Identification:} Lines surpassing the specified probability thresholds were flagged as potential dark patterns, with their corresponding categories identified, enabling detection and categorization of dark patterns on e-commerce websites.

\section{RESULTS}

\begin{table}[htbp]
    \centering
    \caption{Dark Pattern Detection Results}
    \label{tab:results}
    \begin{tabular}{lcc}
        \toprule
        \textbf{Parameters} & \textbf{website1} & \textbf{website2} \\
        \midrule
        Forced Action       & 0               & 0             \\
        Misdirection        & 0.06            & 0.1           \\
        Not Dark Pattern    & 0.75            & 0.68          \\
        Obstruction         & 0               & 0             \\
        Scarcity            & 0.2             & 0.02          \\
        Sneaking            & 0               & 0             \\
        Social Proof        & 0.07            & 0.11          \\
        Urgency             & 0.073           & 0.002         \\
        \bottomrule
    \end{tabular}
\end{table}

    1. After running the extension on both website1 and website2 websites, it is evident that website1 provides better results than website2. The Not Dark Pattern value for website1 is 0.75, which is higher than website2's value of 0.68. This indicates that website1 has fewer dark patterns than website2.
    
    2. When comparing other parameters, we can see that website2 provides better results for scarcity, with a value of 0.02, compared to website1's value of 0.2. This means that website2 is not pressuring users to make quick decisions or purchase products.
    
    3. Next, we can compare the social proof parameter, where website1 has a value of 0.07, while website2 has a value of 0.11. This indicates that website1 is not using social peer pressure to influence user behavior to buy their products, while website2 is slightly better at this.
    
    4. When comparing the urgency parameter, website1 has a value of 0.073, while website2 has a value of 0.002. Here, website2 provides better results as it does not pressure users by showing a false sense of urgency or limited time period offers to act quickly without careful consideration. website1 is seen pressuring users by showing more limited-time offers than website2, which can be considered a dark pattern.
    
    5. Finally, comparing the misdirection parameter, website1 has a value of 0.06, while website2 has a value of 0.1. This means that website1 is not misdirecting users to buy their products by creating false assumptions.

\begin{figure}
    \centering
    \begin{tikzpicture}
        \node[inner sep=0pt] (image) at (0,0) {\includegraphics[width=1\linewidth]{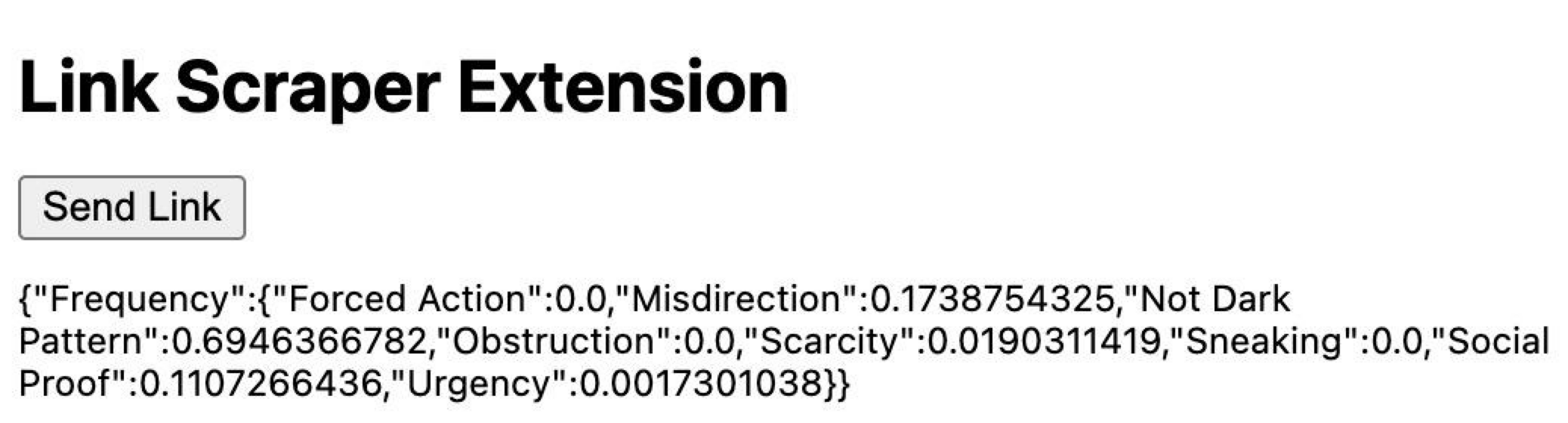}};
        \draw[black, line width=1.2pt] (image.south west) rectangle (image.north east);
    \end{tikzpicture}
    \caption{Famous e-commerce website 1 }
\end{figure}

\begin{figure}
    \centering
    \begin{tikzpicture}
        \node[inner sep=0pt] (image) at (0,0) {\includegraphics[width=1\linewidth]{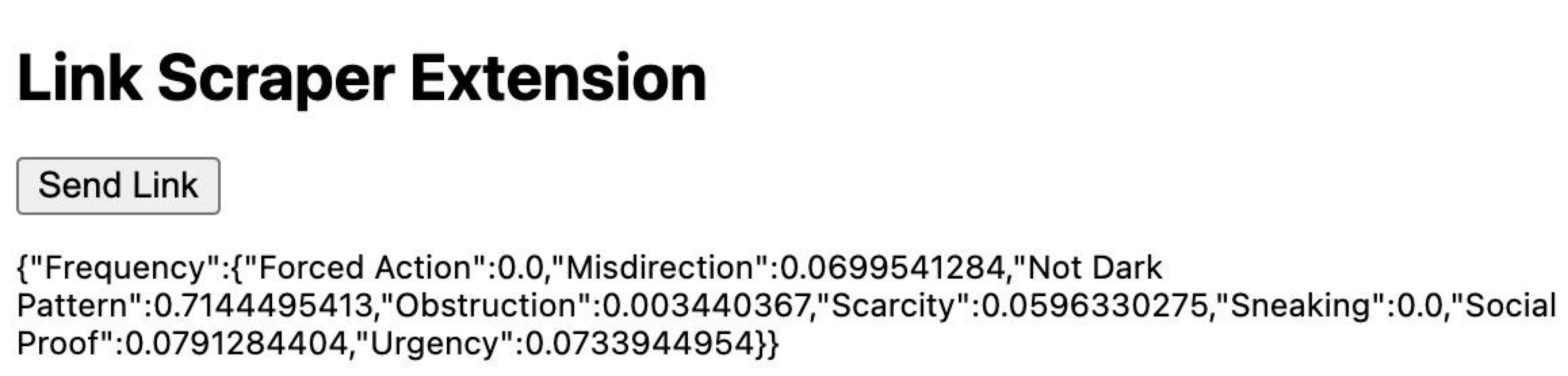}};
        \draw[black, line width=1.2pt] (image.south west) rectangle (image.north east);
    \end{tikzpicture}
    \caption{Famous e-commerce website 2 }
\end{figure}

\vspace{5mm}

\textit{BERT proves to be more accurate among these plus has more contextual understanding (generative ability).}

\begin{table}[htbp]
  \centering
  \caption{Accuracy of Different Models}
  \label{tab:accuracy}
  \begin{tabular}{lcccc}
    \toprule
    & BERT & RNN & GBT & LR \\
    \midrule
    Accuracy (\%) & 96 & 67 & 89 & 91 \\
    \bottomrule
  \end{tabular}
\end{table}

\section{MODEL RESULTS}

\begin{table}[h]
\centering
\caption{Training Results : Accuracy: 0.9597}
\begin{tabular}{|c|c|c|c|}
\hline
\textbf{Epoch} & \textbf{Loss} & \textbf{Accuracy} & \textbf{Validation Accuracy} \\ \hline
1              & 1.2454        & 0.6226             & 0.9322                       \\ \hline
2              & 0.4275        & 0.9363             & 0.9492                       \\ \hline
3              & 0.1998        & 0.9655             & 0.9534                       \\ \hline
4              & 0.1221        & -                  & 0.9513                       \\ \hline
5              & 0.0822        & 0.9867             & 0.9597                       \\ \hline
\end{tabular}
\label{tab:training_results}
\end{table}

\begin{table}[htbp]
    \centering
    \caption{Not a dark pattern}
    \begin{tabular}{|>{\centering\arraybackslash}m{0.3\linewidth}|>{\centering\arraybackslash}m{0.6\linewidth}|}
        \hline
        \textbf{Statement} & "me and my friends are going to buy shoes which are 20\% off" \\
        \hline
        \textbf{Category} & \textbf{Probability (\%)} \\
        \hline
        Forced Action & 1.74 \\
        Misdirection & 5.83 \\
        Not Dark Pattern & 54.21 \\
        Obstruction & 1.43 \\
        Scarcity & 0.52 \\
        Sneaking & 2.75 \\
        Social Proof & 31.51 \\
        Urgency & 2.01 \\
        \hline
        \multicolumn{2}{|c|}{\textbf{Predicted Category: Not Dark Pattern}} \\
        \hline
    \end{tabular}
\end{table}
\vspace{5mm}
\textit{The model effectively identifies and filters out dark patterns through its contextual comprehension. It demonstrates reliable performance in distinguishing deceptive design elements, enhancing user protection.}

\vspace{5mm}

\textit{By leveraging its understanding of contextual cues and generative ability, the model helps mitigate the impact of dark patterns, fostering a safer and more transparent online environment.}

\begin{table}
    \centering
    \caption{Dark Pattern Analysis}
    \begin{tabular}{|>{\centering\arraybackslash}m{0.2\linewidth}|>{\centering\arraybackslash}m{0.6\linewidth}|}
        \hline
        \textbf{Statement} & \hl{"Hurry! Only 2 left in stock"} \\
        \hline
        \textbf{Category} & \textbf{Probability (\%)} \\
        \hline
        Forced Action & 0.23 \\
        Misdirection & 0.29 \\
        Not Dark Pattern & 0.24 \\
        Obstruction & 0.29 \\
        Scarcity & 98.09 \\
        Sneaking & 0.24 \\
        Social Proof & 0.26 \\
        Urgency & 0.35 \\
        \hline
        \multicolumn{2}{|c|}{\textbf{Predicted Category: Scarcity}} \\
        \hline
    \end{tabular}
    \centering
    \caption{Dark Pattern between two non-meaningful sentences}
    \begin{tabular}{|>{\centering\arraybackslash}m{0.25\linewidth}|>{\centering\arraybackslash}m{0.6\linewidth}|}
        \hline
        \textbf{Statement} & "sdjbfksbdfgbkldsglkdflgf \hl{subscribe now or regret the offer of 20\%} djkbfksjbglsbdfsdbfksdbfgkjsbdkgbskdbfsdbfsd" \\
        \hline
        \textbf{Category} & \textbf{Probability (\%)} \\
        \hline
        Forced Action & 2.07 \\
        Misdirection & 47.84 \\
        Not Dark Pattern & 39.92 \\
        Obstruction & 3.90 \\
        Scarcity & 0.49 \\
        Sneaking & 2.92 \\
        Social Proof & 0.94 \\
        Urgency & 1.92 \\
        \hline
        \multicolumn{2}{|c|}{\textbf{Predicted Category: Misdirection}} \\
        \hline
    \end{tabular}
    \centering
    \caption{Dark Pattern between two meaningful sentences}
    \begin{tabular}{|>{\centering\arraybackslash}m{0.25\linewidth}|>{\centering\arraybackslash}m{0.6\linewidth}|}
        \hline
        \textbf{Statement} & "My name is Jin Kazama and I am in Pune, \hl{get 30\% off on this bottle but you'll have to sign up first or you'll miss it}, let's go have camping together" \\
        \hline
        \textbf{Category} & \textbf{Probability (\%)} \\
        \hline
        Forced Action & 2.01 \\
        Misdirection & 69.15 \\
        Not Dark Pattern & 20.86 \\
        Obstruction & 2.60 \\
        Scarcity & 0.86 \\
        Sneaking & 2.14 \\
        Social Proof & 1.66 \\
        Urgency & 0.73 \\
        \hline
        \multicolumn{2}{|c|}{\textbf{Predicted Category: Misdirection}} \\
        \hline
    \end{tabular}
    
\end{table}




\vspace{10mm}

\section{Literature Survey}

[1] The paper introduces a mobile app solution that aims to detect and alert users about deceptive design elements encountered during browsing, with a focus on achieving accurate and high-performance detection.\vspace{2mm}

[2] The paper presents UIGuard, a knowledge-driven system that utilizes computer vision and natural language pattern matching to automatically identify a wide range of dark patterns in mobile user interfaces (UIs). The system demonstrates superior performance, achieving a precision of 0.82, recall of 0.77, and an F1 score of 0.79 in dark pattern detection.\vspace{2mm}

[3] The study examines user perceptions of a Chrome extension called the Dark Pattern Detector, which aims to identify and inform users about dark patterns. The study finds that users respond positively to the extension and provides valuable feedback for further improvement.\vspace{2mm}

[4] The study investigates the presence of dark patterns in 240 mobile apps, revealing that 95\% contain deceptive interfaces. It also conducts an online experiment with 589 users, demonstrating that users can better recognize malicious designs when informed about the issue of dark patterns.\vspace{2mm}

[5] The paper introduces AIDUI, an automated approach that combines computer vision and natural language processing to detect and classify ten unique UI dark patterns. The system achieves an overall precision of 0.66, recall of 0.67, and an F1-score of 0.65, with potential for future improvements in detecting additional patterns.\vspace{2mm}

[6] The paper analyzes 300 data collection consent notices from news outlets to uncover strategies, termed "dark patterns," that circumvent GDPR compliance, thereby highlighting the challenges in ensuring informed consent in online services.\vspace{2mm}

[7] The study by Mathur et al. (2019) examines dark patterns on a large scale by crawling 11,000 shopping websites and identifies various types of dark patterns observed.\vspace{2mm}

[8] The paper by Bhoot et al. (2020) analyzes end-user reactions and perspectives toward the identification of dark patterns, contributing to the understanding of user awareness and perceptions of these deceptive practices.\vspace{2mm}

[9] The paper by Curley et al. (2021) proposes a framework for detecting web-based dark patterns, aiming to shed light on these deceptive practices and facilitate their identification and mitigation.\vspace{2mm}

[10] The study by Hausner and Gertz (2021) investigates the presence of dark patterns in the interaction with cookie banners, highlighting the deceptive tactics employed in these consent mechanisms.\vspace{2mm}

[11] The paper by Soe et al. (2022) explores automated methods for detecting dark patterns in cookie banners, discussing the challenges and limitations of this task and proposing potential approaches.\vspace{2mm}

[12] The paper by Bösch et al. (2016) discusses privacy dark strategies and patterns, examining the various techniques and tactics employed to manipulate user privacy choices and decisions.\vspace{2mm}

[13] The study by Bongard-Blanchy et al. (2021) examines dark patterns from the end-user perspective, providing insights into how users perceive and experience these deceptive practices.\vspace{2mm}

[14] The paper by Luguri and Strahilevitz (2024) sheds light on the concept of dark patterns, exploring their implications and potential consequences for users and society.\vspace{2mm}

[15] The study by Gray et al. (2023) contributes towards developing a preliminary ontology of dark patterns knowledge, aiming to systematize and organize the understanding of these deceptive practices.\vspace{2mm}

[16] The work by Martini et al. (2021) explores the phenomenology of dark patterns and examines the legal responses and potential regulatory measures to address these deceptive practices.\vspace{2mm}

[17] The systematic literature review conducted by Gray et al. (2023) maps the landscape of dark patterns scholarship, providing a comprehensive overview of the research in this field.\vspace{2mm}

[18] The paper by Yada et al. (2022) presents a dataset and baseline evaluations of dark patterns in e-commerce settings, contributing to the development of automated detection methods and the understanding of these practices in online shopping environments.\vspace{2mm}

[19] The comparative study by Gunawan et al. (2021) examines dark patterns across mobile and web platforms, highlighting the similarities and differences in the manifestation of these deceptive practices across different modalities.\vspace{2mm}

[20] The paper by Gray et al. (2021) evaluates dark patterns in consent banners from an interaction criticism perspective, analyzing the legal requirements and implications of these deceptive practices in the context of obtaining user consent.

\section{CONCLUSION}
In conclusion, the evaluation of dark pattern detection results reveals valuable insights into the user experience strategies employed by e-commerce platforms. While some platforms demonstrate stronger adherence to ethical design principles, others exhibit tendencies towards employing manipulative tactics.

The analysis underscores the importance of continuously monitoring and evaluating user interfaces to identify and mitigate the presence of dark patterns. By prioritizing transparency, clarity, and user empowerment, businesses can foster trust and loyalty among their customer base. Furthermore, regulatory bodies and industry stakeholders play a crucial role in setting standards and enforcing ethical practices to safeguard consumer interests in the digital marketplace.

\section{REFERENCE}

[1] Dr. S. Hrushikesava Raju et al. (2022) shed light on enhancing user awareness regarding deceptive practices through app interfaces.

[2] Chen, J., Sun, J., Feng, S., Xing, Z., Lu, Q., \& Xu, X. (2023) present an automated approach to identify deceptive design techniques in mobile applications.

[3] Wood, R., Brown, D. C., Dunlap, D. R., \& McCrickard, D. S. (May 8, 2023) assess the impact of detecting deceptive design on online user behavior.

[4] Di Geronimo, L., Braz, L., Fregnan, E., \& Bacchelli, A. (April 2020) explore the design of mobile apps and user interpretation.

[5] Mansur, S. M. H., Salma, S., Awofisayo, D., \& Moran, K. (March 12, 2023) introduce AidUI, a system advancing automated detection of deceptive design elements in user interfaces.

[6] Soe, T. H., Nordberg, O. E., Guribye, F., \& Slavkovik, M. (June 24, 2020) examine deceptive design tactics in cookie consent interfaces for online news platforms.

[7] Mathur, A., Acar, G., Friedman, M. J., Lucherini, E., Mayer, J., Chetty, M., \& Narayanan, A. (November 2019) discuss dark patterns observed in a large-scale crawl of shopping websites.

[8] Bhoot, A. M., Shinde, M. A., \& Mishra, W. P. (November 05–08, 2020) analyze end-user reactions towards identifying dark patterns.

[9] Curley, A., O'Sullivan, D., Gordon, D., Tierney, B., \& Stavrakakis, I. (July 2021) propose a framework for detecting web-based dark patterns.

[10] Hausner, P., \& Gertz, M. (March 27, 2021) investigate dark patterns in interactions with cookie banners.

[11] Soe, T. H., Santos, C. T., \& Slavkovik, M. (April 21, 2022) explore automated detection methods for dark patterns in cookie banners.

[12] Bösch, C., Erb, B., Kargl, F., Kopp, H., \& Pfattheicher, S. (June 2, 2016) discuss privacy dark strategies and patterns.

[13] Bongard-Blanchy, K., Rossi, A., Rivas, S., Doublet, S., \& Koenig, V. (June 28–July 2, 2021) examine dark patterns from an end-user perspective.

[14] Luguri, J., \& Strahilevitz, L. J. (April 19, 2024) illuminate dark patterns and their implications.

[15] Gray, C. M., Santos, C., \& Bielova, N. (2023) contribute towards developing an ontology of dark patterns knowledge.

[16] Martini, M., Drews, C., Seeliger, P., \& Weinzierl, Q. (2021) explore the phenomenology and legal responses to dark patterns.

[17] Gray, C. M., Chamorro, L. S., Obi, I., \& Duane, J.-N. (2023) conduct a systematic literature review mapping the landscape of dark patterns scholarship.

[18] Yada, Y., Feng, J., Matsumoto, T., Fukushima, N., Kido, F., \& Yamana, H. (November 12, 2022) present a dataset and baseline evaluations of dark patterns in e-commerce.

[19] Gunawan, J., Pradeep, A., Choffnes, D., Hartzog, W., \& Wilson, C. (October 2021) compare dark patterns across mobile and web platforms.

[20] Gray, C. M., Santos, C., Bielova, N., Toth, M., \& Cliford, D. (May 8–13, 2021) examine dark patterns in consent banners from an interaction criticism perspective.

\end{document}